\documentclass[aps,10pt,twocolumn,floatfix,amsmath,amssymb,pra,showpacs,longbibliography,raggedbottom]{revtex4-2}
\usepackage{graphicx}
\usepackage{color}

\usepackage{bm}
\usepackage{amsmath}

\begin{document}

\title{Eigenvalue sensitivity from eigenstate geometry\\near and beyond arbitrary-order exceptional points}

\author{Henning Schomerus}
\affiliation{Department of Physics, Lancaster University, Lancaster, LA1 4YB, United Kingdom}

\begin{abstract}
Systems with an effectively non-Hermitian Hamiltonian display an enhanced sensitivity to parametric and dynamic perturbations, which arises from the nonorthogonality of their eigenstates. This enhanced sensitivity can be quantified by the phase rigidity, which mathematically corresponds to the eigenvalue condition number, and physically also determines the Petermann factor of quantum noise theory.
I derive an exact nonperturbative expression for this sensitivity measure that applies to arbitrary eigenvalue configurations. The expression separates spectral correlations from additional geometric data, and
retains a simple asymptotic behaviour close to exceptional points (EPs) of any order, while capturing the role of additional states in the system. This reveals that such states can have a sizable effect even if they are spectrally well separated, and identifies the specific matrix whose elements determine this nonperturbative effect.
The employed algebraic approach, which follows the eigenvectors-from-eigenvalues school of thought, also provides direct insights into the geometry of the states near an EP. For instance, it can be used to show that the phase rigidity follows a striking equipartition principle in the quasi-degenerate subspace of a system.
\end{abstract}

\maketitle

\section{Introduction}

From quantum mechanics to classical physics, linear algebra has become such a central tool for the description of physical systems that it seems to hardly hold any new surprises.
Experience informs us about the utility of eigenvalues and eigenstates, whose calculation appears to be challenging only when one has to deal with very large systems.
An interesting mathematical complication arises in effective descriptions that result in non-orthogonal eigenstates, as is common in open quantum systems and classical systems with gain and loss, where the effective Hamiltonian becomes non-Hermitian \cite{moiseyev2011non,ashida2020}. A full characterization of each state then has to involve two variants of it, the right eigenstate $|R_i\rangle$ and the left eigenstate $\langle L_i|$ of any eigenvalue $E_i$. As long as eigenvalues are nondegenerate, these two sets can be used to form a biorthogonal basis that diagonalizes the system  \cite{kato2013perturbation}.
However, the nonorthogonality becomes further accentuated at generic eigenvalue degeneracies, so-called exceptional points (EPs), which lead to a defective system that cannot be diagonalized anymore  \cite{kato2013perturbation,Heiss2000,Berry2004,Heiss2004}. A physical signature of these EPs is a drastically altered sensitivity of the system to perturbations, both statically \cite{Wie14,Wiersig20} as well as dynamically  \cite{Sch20,Hashemi2022,Budich2020}. The non-orthogonality itself  enhances this sensitivity already for spectrally isolated states, which can be quantified  by a number of equivalent quantities, such as the phase rigidity
\begin{equation}
\label{eq:r}
r_i=\frac{\langle L_i|R_i\rangle}{\sqrt{\langle L_i|L_i\rangle\langle R_i|R_i\rangle}},
\end{equation}
which naturally appears in the characterization of the flux nonconservation  \cite{vanLangen1997}, or the Petermann factor
\begin{equation}
K_i=|r_i|^{-2},
\end{equation}
which naturally appears in the evaluation of quantum noise \cite{Petermann79,Siegman89,Patra2000,Berry2003,Yoo11,Simonson22}, as well as in classical dynamical response theory \cite{Sch20,Hashemi2022}.
Approaching an EP, the phase rigidity  tends to zero, and the diverging Petermann factor signifies the abovementioned drastic change of the response to perturbations.
Experimentally, these effects have been observed, e.g., in the broadening of the quantum-limited linewidth of lasers away from EPs \cite{vanEijkelenborg1996,Cheng1996}, in the change of their lineshape at EPs \cite{Takata21}, and in the enhancement of both signal and noise in phonon lasers \cite{Zhang2018} and laser gyroscopes \cite{Wang2020}.
In non-reciprocal settings, the Petermann factor can also diverge at phase transitions that localize the right and left eigenstates of a given mode at opposite ends of a system, which then results in a proposed transition to directed amplification and sensing \cite{Sch20,Budich2020,Wanjura2020,McDonald2020}.
As non-Hermitian spectra furthermore lead to highly complex structures in parameter space \cite{Doppler2016,Ber19,Kaw19}, a universal and detailed characterization is challenging, which singles them out as an active field of study.

On the mathematical side, the very same quantity $r_i$ is know as the eigenvalue condition number, which features practically, e.g., as a measure of accuracy of numerical diagonalization algorithms \cite{trefethen2005spectra}.
The concrete expression \eqref{eq:r} then reveals a deep relation between eigenvectors and spectral properties---even though the latter are, in principle, basis invariant. These correlations are born out, e.g., in the study of generic non-Hermitian random matrices \cite{Chalker1998}, and in
causality  constraints of passive systems \cite{Wiersig2019,Schomerus2022,Wiersig2022}.
In  separate developments, even more concrete eigenvector-eigenvalue relations are receiving considerable mathematical attention. This follows the realization that such relations can provide deeply surprising and general insights, even into normal systems with orthogonal eigenvectors. While variants of such relations have appeared in many specific contexts, their comprehensive landscape was only fully appreciated very recently, as is beautifully surveyed in Ref.~\cite{denton2022eigenvectors}. As this reference hints at various points, the underlying ideas also transfer to non-normal systems, which is the general mathematical feature that renders eigenvectors non-orthogonal.

In this work I combine both themes to provide an exact nonperturbative reformulation of the phase rigidity which applies to arbitrary eigenvalue configurations, and separates spectral information that becomes singular at EPs from geometric data that remains regular.
This reformulation allows us to extract valuable information about the properties of a system both close to as well as away from EPs of arbitrary order, including combinations  where some of the eigenvalues are quasi-degenerate and others are not, but the additional eigenstates still have a finite overlap with the quasi-degenerate subspace.
This reveals that such additional states can have a direct effect even if they are spectrally well separated, and captures this precisely---realizing the \emph{beyond} part in the title.
On the other hand, in the limit of a system where all states participate in the EP, we recover a recently derived compact asymptotic result \cite{Wiersig2023}---realizing the \emph{near} part in the title.
Furthermore, the underlying algebraic features allow us to extract additional geometric information about the states near the EP, revealing a striking equipartition property of the contributions from the different directions in the quasi-degenerate subspace.
These features  directly transfer to the  response of physical systems to  parametric and dynamic perturbations, including quantum and classical noise, and determine the function of these systems as sensors, in the settings described further above.

These results are presented along the following lines.
Section \ref{sec:background} provides the mathematical and theoretical background for this work.
Section \ref{sec:result} formulates and derives the exact nonperturbative
reformulation of the phase rigidity, which constitutes
the main general result of this work.
Section \ref{sec:asymptotic} highlights the implications for asymptotics near EPs, which constitutes the main application of this formalism.
Sec.~\ref{sec:proof2} illuminates these features further by an explicitly constructive approach to the eigenstate geometry, while Sec.~\ref{sec:equi}
 formulates the equipartition principle that arises from these considerations.
Section \ref{sec:examples} provides two  examples illustrating the nonperturbative role of additional states and the interplay of multiple EPs, and Sec.~\ref{sec:conclusions} contains the conclusions.

\section{Mathematical background}\label{sec:background}
In this paper, we are interested in geometric properties of eigenvectors of non-Hermitian matrices, as captured by the phase rigidity $r_i$ given in Eq.~\eqref{eq:r}. In this section, I introduce the central concepts surrounding this quantity in the notation common to the physics literature, which is mainly concerned with effective Hamiltonians $H$ and employs the Dirac notation.
Furthermore, I review the starting point of the considerations, given by the asymptotic results of the phase rigidity near maximally degenerate EPs in Ref.~\cite{Wiersig2023}, and introduce the mathematical objects that we need to formulate, derive, and apply the main result of the present work, which extends this framework to systems with arbitrary eigenvalue configurations.

\subsection{Exceptional points and phase rigidity}
In the physical context of effective Hamiltonians $H$,  the specific eigenvalues determine resonance energies or frequencies, which I will denote by the symbol  $E_i$.
These eigenvalues are defined via the right and left eigenvalue problems as
\begin{align}
H|R_i\rangle=E_i|R_i\rangle,\qquad
\langle L_i| H=\langle L_i|E_i,
\label{eigval}
\end{align}
where $|R_i\rangle$, $\langle L_i|$ are the corresponding right and left eigenvectors.
Throughout, we will deal with systems of finite dimension $m=\mathrm{dim}(H)$, so that the eigenvalue spectrum is discrete.
Biorthogonality implies that $\langle L_i|R_j\rangle=0$ if $E_i\neq E_j$. For reasons that we explain next, we will not enforce the biorthogonal normalization condition
$\langle L_i|R_i\rangle=1$. Instead, we will present results in a form that is independent of this normalization condition, or, where convenient for the exposition, resort to the unproblematic choice of individually normalised eigenvectors $\langle \hat R_i| \hat R_i\rangle=\langle \hat L_i| \hat L_i\rangle=1$ (the hat then clearly indicates this choice).

These considerations are enforced by the behaviour of these states near degeneracies.
For degenerate eigenvalues, we have to distinguish their algebraic multiplicity $n$ and geometric multiplicity $d$, where $n$ is given by the multiplicity of the root $p(E_i)=0$ of the characteristic polynomial
\begin{equation}
\label{eq:cpol}
p(E)=\mathrm{det}(E\openone-H),
\end{equation}
and $d$ by the number of linearly independent solutions in each of the two eigenvalue equations \eqref{eigval} (here and henceforth,
$E$ represents a continuous energy or frequency variable).
In the case of normal matrices, which include Hermitian and unitary ones, both multiplicities $d=n$ coincide, while in the case of non-normal ones,
generically $d=1$, hence only a single eigenvector can be found. These degeneracies are known as EPs of order $n$, and they imply that  the matrix can no longer be diagonalized--in other words, the eigensystem is defective.
(Instead, by  Schur's unitary triangularization theorem the effective Hamitonian can always be unitarily transformed into upper-triangular form, for which we then explicitly use the symbol $T$ \footnote{Utilizing a similarity transformation, at an EP the system may further be brought into a Jordan normal form. However, this removes the  geometric relations captured by $r_i$, whose physical content is fixed by the scalar product in the original Hilbert space.}.)

The quantities at the center of this paper provide insight into the geometric properties of the eigensystem, both for general fixed eigenvalue configurations as well as for the dependence in parameter space.
In the parametric vicinity of the EP, the degeneracy of the eigenvalues in question is generically completely lifted. As one approaches the EP, not only $n$ eigenvalues tend to a common limit $E_{EP}$, but also the corresponding eigenvectors approach common limits $|R_{EP}\rangle$ and $\langle L_{EP}|$  (modulo normalization and phase, which can be suitably fixed). On the other hand, right up to the point of degeneracy the still-nondegenerate eigenvectors fulfill the biorthogonality condition $\langle L_i|R_j\rangle=0$ for $i\neq j$.
This implies that at the EP, the right and left eigenvectors become self-orthogonal,  $\langle L_{EP}|R_{EP}\rangle=0$.
The phase rigidity $r_i$ from Eq.~\eqref{eq:r} quantifies this self-overlap at any distance away from the EP, and indeed does so independent of the normalization choice mentioned above (it however contains an arbitrary phase). In terms of  the individually normalized eigenstates, it can also be written as
\begin{equation}
\label{eq:normalised}
r_i=\langle \hat L_i | \hat R_i\rangle,
\end{equation}
hence, reduces to the overlap of these states.

\subsection{Phase rigidity near EPs of maximal order}
From Ref.~\cite{Wiersig2023}, the behaviour of $r_i$ is known near maximally degenerate EPs, where all eigenvalues and eigenstates of the system approach each other, i.e., for the case where the order of the EP equals the dimension of the effective Hamiltonian, $n=m$.
Near these EPs, the Hamiltonian can be written as
\begin{equation}\label{eq:h}
H=E_{EP}\openone    +N+\varepsilon H',
\end{equation}
where $\varepsilon  H'$ denotes a perturbation, which is assumed to be small, while the residual term $N$ obeys
\begin{equation}
N^{n-1}\neq 0,\quad N^n=0.
\end{equation}
For this setting, Ref.~\cite{Wiersig2023} derives the compact asymptotic expression
\begin{equation}
\label{eq:jan}
|r_i|\sim|n(E_i-E_{EP})^{n-1}/\xi|
\end{equation}
for the phase rigidity of the individual states near the EP, where
\begin{equation}
\xi=||N^{n-1}||_2
\end{equation}
is a common spectral strength characterizing the EP itself
(we use $\sim$ to denote the leading-order asymptotic result for $\varepsilon\to 0$, including coefficients). Involving just the perturbed eigenvalue and a single characteristic number that can be evaluated directly at the EP, this relation beautifully accentuates the spectral significance of this quantity.
On the other hand, the result is only valid close to maximally degenerate EPs in the space of $H$, while in the case that this effective Hamiltonian is obtained by a truncation to a quasi-degenerate subspace the quantity $\xi$ does not capture contributions from additional states that do not participate in the EP.

\subsection{Eigenvalues from eigenvectors}
The main goal of this work is to express the phase rigidity in a general and universal form so that it also applies in presence of additional states that do not participate in the EP, and indeed for any eigenvalue configuration. The methods are inspired from the eigenvalue-from-eigenvector school of thought, which has been fully developed for Hermitian systems, where we can identify $|R_i\rangle=|L_i\rangle$  \cite{denton2022eigenvectors}.
In the notation of the present paper, the components of these conventional eigenvectors then obey, after normalization, the identity
\begin{equation}
\label{eq:evalvec}
|\langle k|\hat R_i\rangle|^2=|\langle \hat  L_i | k \rangle|^2=\frac{\prod_{j=1}^{m-1} (E_i-\tilde E_{j,k})}{\prod_{j=1; j\neq i}^{m} (E_i-E_j)},
\end{equation}
where $|k\rangle$  denotes a fixed orthonormal basis,
while $\tilde E_{j,k}$ are the eigenvalues of the matrix $\tilde H_k$ that is obtained from $H$ by removing its $k$th row and column.

In Ref.~\cite{denton2022eigenvectors}, these truncated matrices are referred to as \emph{minors}. In the present work, we follow the more standard convention to reserve this word for \emph{determinants} of such truncated matrices. In particular, we will encounter determinants of matrices obtained by removing a row $k$ and a column $l$ from a matrix $A$, and denote these minors as $M_{kl}(A)$. We furthermore will encounter the adjugate matrix $\mathrm{adj}(A)$, whose matrix elements are given by
\begin{equation}
\label{eq:adj}
\langle l|\mathrm{adj}(A) |k\rangle=(-1)^{kl}M_{kl}(A).
\end{equation}
As we will see, such matrix elements efficiently capture the nontrivial geometric properties of eigenvectors in non-Hermitian settings.

\subsection{The role of the characteristic polynomial}
\label{sec:cp}
The final piece of mathematical preparation concerns the role of characteristic polynomial in linking these previous statements.
We see that this polynomial features naturally both in Eq.~\eqref{eq:jan}, which we  aim to generalize, and  Eq.~\eqref{eq:evalvec}, which relates to the general tools required to formulate and derive this generalization.

Generally, this polynomial can be written in two explicit forms,
\begin{equation}
p(E)=\sum_{k=0}^m E^k a_{m-k}=\prod_{j=1}^m(E-E_j),
\end{equation}
where in the second expression any degenerate eigenvalues are repeated with their algebraic multiplicity.

Let us first establish the asymptotics of this polynomial near EPs of maximal order $n=m$, as modeled by the Hamiltonian given in Eq.~\eqref{eq:h}.  To simplify the derivation, I set $E_{EP}=0$, and reinstate it to a finite value at the end of this section. We then can use Newton's formulas \cite{Haake1996} to express the coefficients $a_k$ by the traces $t_k=\mathrm{tr}\,H^k$, which here all are of  order $t_k=O(\varepsilon)$ (the EP condition implies  $\mathrm{tr}\,N^k=0$ for all $k\geq 1$.)
With our choice $E_{EP}=0$,  we then find that with the exception of $a_0=1$, each coefficient $|a_k|\sim |t_k/k|$ is of the same order, and by comparing orders we only need to consider
\begin{equation}
\label{eq:pol}
p(E)\sim E^n +(-1)^n \mathrm{det}(H).
\end{equation}
Therefore, close to the EP the eigenvalues approximately spread out to take uniformly spaced positions on a circle, which recovers the well-know result of the eigenvalue cloud close to the EP \cite{kato2013perturbation}. In the present context, Eq.~\eqref{eq:pol}
allows us to immediately read off $p'(E_i)\sim n E_i^{n-1}$,
where the prime denotes the derivative of the polynomial with respect to $E$.
Reinstating $E_{EP}$ to a finite value, this turns into the asymptotic expression
\begin{equation}
\label{eq:pprime}
p'(E_i)\sim n (E_i-E_{EP})^{n-1}.
\end{equation}
We see that this captures the spectral content of Eq.~\eqref{eq:jan}.
In the presence of additional states, we analogously infer
\begin{equation}
\label{eq:cpasym}
p'(E_i)\sim n (E_i-E_{EP})^{n-1}\prod_k'(E_k-E_{EP}),
\end{equation}
where the product is over the remaining eigenvalues not participating in the EP (using again their algebraic multiplicities in case that they themselves participate in their own EPs).

Finally, we remark that irrespective of the eigenvalue configuration, the result \eqref{eq:evalvec} for eigenvectors of Hermitian matrices can be written as
\begin{equation}
\label{eq:evalvecs}
|\langle k|\hat R_i\rangle|^2=|\langle \hat  L_i | k \rangle|^2=\frac{\tilde p_k(E_i)}{p'(E_i)},
\end{equation}
where $\tilde p_k(E)$ is the characteristic polynomial of the matrix $\tilde H_k$  (i.e., $H$ with its row and column $k$ removed).

In summary, this background highlights the relation of eigenstate geometry to algebraic objects such as the characteristic polynomial \eqref{eq:cpol} and the adjugate matrix \eqref{eq:adj}.
In the next section, we use these mathematical objects to formulate and derive the main general result of the present work.

\section{Main general result}\label{sec:result}
With these mathematical preparations, we can now formulate the main general result of this work, a nonperturbative expression of the phase rigidity that applies to arbitrary eigenvalue configurations. In this section, I first state this result, and then turn to its derivation, which usefully illuminates general geometric relations of the involved states. Subsequent sections then describe the concrete insights that one can gain from this reformulation, such as the ensuing general asymptotic behaviour near EPs, including for the case where these do not involve all states in the system. This utilizes a second key feature of the intended reformulation, namely, that it cleanly separates contributions due to spectral correlations, which can be analyzed perturbatively, from additional, nonperturbative, geometric data.

\subsection{Exact reformulation}
To achieve these goals, I proof, below in this section, the \emph{exact} reformulation
\begin{subequations}\label{eq:result0}
\begin{equation}
\label{eq:result0a}
r_i=\frac{p'(E_i)}{g_i}
\end{equation}
of the phase rigidity,
where the prime again denotes the derivative of the characteristic polynomial $p(E)$ with respect to $E$,
while
\begin{equation}
\label{eq:adef}
g_i=\langle \hat R_i|\mathrm{adj}(E_i\openone-H)|\hat L_i\rangle
\end{equation}
\end{subequations}
determines a matrix element of the adjugate matrix, obtained from the determinantal minors $M_{kl}(E_i\openone-H)$ as specified in Eq.~\eqref{eq:adj}.

This reformulation allows to obtain the phase rigidity of eigenvalues for arbitrary configurations, both close and far away from EPs, and irrespective of how many of the eigenvalues in the system are participating in it.
Furthermore, it cleanly separates out spectral content via the characteristic polynomial, while additional geometric data is captured in the coefficient $g_i$.

\subsection{Resolvent proof}
\label{sec:proof}

For the derivation we employ the resolvent method, which features centrally in the eigenvectors-from-eigenvalue context \cite{denton2022eigenvectors}.
On the physical side, this approach mirrors closely contexts involving the Petermann factor $K_i=1/|r_i|^2$. This factor typically appears from the resolvent
\begin{equation}
G=(E\openone -H)^{-1},
\end{equation}
which we can analyze for $E\to E_i$.
Using the spectral decomposition, we  have
\begin{equation}
G_{kl}\sim \frac{R_{i,k}L_{i,l}^*}{\langle L_i|R_i\rangle}\frac{1}{E-E_i},
\end{equation}
where $\sim$ now also implies the asymptotics $E\to E_i$.
On the other hand, using Cramer's rule,
\begin{equation}
G_{kl}\sim \frac{(-1)^{k+l}M_{kl}(E_i\openone-H)}{p'(E_i)}\frac{1}{E-E_i},
\end{equation}
where $M_{kl}(A)$ is the minor of $A$ obtained by removing the indicated row-column pair, and the prime in $p'(E_i)$ denotes the derivative of $p(E)$ with respect to $E$, just as introduced in the background Sec.~\ref{sec:background}.

Comparison of these two expressions gives the important exact relation  (see remark 5 in Ref.~\cite{denton2022eigenvectors})
\begin{equation}
\label{eq:imp}
(-1)^{k+l} M_{lk}(E_i\openone-H)=\frac{R_{i,k}L_{i,l}^*}{\langle L_i|R_i\rangle}p'(E_i).
\end{equation}
We now temporarily resort to the individually normalized eigenvectors, $\langle \hat R_i| \hat R_i\rangle=1=\langle \hat  L_i| \hat L_i\rangle$
to resolve this into the relation
\begin{equation}
\langle \hat  R_i|\mathrm{adj}(E_i\openone-H)| \hat  L_i\rangle = \frac{p'(E_i)}{\langle \hat  L_i| \hat R_i\rangle}
\end{equation}
for the corresponding matrix element of the adjugate matrix.
Given this normalization, we immediately obtain the general main result \eqref{eq:result0} via
Eq.~\eqref{eq:normalised}.

\section{Asymptotic behaviour near EPs}
\label{sec:asymptotic}

One of the key merits of the exact reformulation \eqref{eq:result0} is its regular behaviour close to EPs. The characteristic polynomial $p(E)$ factors out the appropriate power-law dependence in $(E_i-E_{EP})$, while the geometric factor $g_i$ remains finite. Furthermore, both factors  fully account for all remaining states in the system, again irrespective of how complicated their eigenvalue configuration may be.  Therefore, we can evaluate the resulting asymptotic behaviour of the phase rigidity not only near EPs of maximal order $n=m$, but also for EPs of any order $n<m$, where additional states exist in the system.

\subsection{EPs of maximal order}

First, we verify that this approach recovers the correct asymptotics near EPs of maximal order, as described by the effective Hamiltonian \eqref{eq:h}.
The exact reformulation  then expresses the result \eqref{eq:jan} as
\begin{equation}
\label{eq:result}
|r_i|\sim\frac{|p'(E_i)|}{|M_{\hat L_{EP}\hat R_{EP}}(N)|},
\end{equation}
where $M_{\hat L_{EP}\hat R_{EP}}(N)$  denotes the minor of $N$ obtained by removing the eigenvector directions $\langle \hat L_{EP}|$ and $|\hat R_{EP}\rangle$ at the EP from the row and column. This direct formulation rests on the fact that these directions are orthogonal, $\langle \hat L_{EP}| \hat R_{EP}\rangle=0$, so that we can make them part of an orthonormal basis.
 For this specific case, the two results agree by the following relations: As already discussed in the context of Eq.~\eqref{eq:pprime}, $p'(E_i)$ exactly captures the spectral information appearing in the numerator of Eq.~\eqref{eq:jan}.
Furthermore, we now can confirm that
\begin{equation}
\xi= |M_{\hat L_{EP}\hat R_{EP}}(N)|
\end{equation}
follows using the normal form where $N$ takes an upper triangular form $T$, as obtained by a Schur decomposition, resulting in a basis change after which $|\hat R_{EP}\rangle$ is the first basis vector and $\langle \hat L_{EP}|$ is the last.
Then we can verify explicitly that
\begin{equation}
\label{eq:denom}
|M_{n1}(T)|=\left|\prod_{k=1}^{n-1} T_{k,k+1}\right|=||T^{n-1}||_2=\xi
\end{equation}
takes exactly the same algebraic form as  obtained from  its definition above; and this then remains true in any orthonormal basis.
Equation \eqref{eq:result} asks us to take the ratio of
Eqs.~\eqref{eq:pprime} and \eqref{eq:denom}, and this then indeed recovers Eq.~\eqref{eq:jan}.

\subsection{EPs of any order}
Secondly, we make use of the universality of Eq.~\eqref{eq:result0} to address the role of additional states in the system, hence, determine the asymptotic behaviour of the phase rigidity near generic EPs to  any order.

Indeed, the derivation of Eq.~\eqref{eq:result0} does not rely on a truncation of the matrix to the quasi-degenerate space, which is required to define $\xi$ in the original context. Therefore, we can replace $N\to H-E_{EP}\openone$ by the full Hamiltonian at the EP, including possible additional states with non-degenerate eigenvalues. This gives our general result for the asymptotic behaviour near such an EP,
\begin{equation}
\label{eq:result2}
|r_i|\sim\frac{|p'(E_i)|}{|M_{\hat L_{EP}\hat R_{EP}}(E_{EP}\openone-H)|},
\end{equation}
where the minor can be calculated directly in an orthonormal basis where $|\hat R_{EP}\rangle$ and $|\hat L_{EP}\rangle$ are basis vectors, utilizing again that such a basis exists since at the EP $\langle \hat L_{EP} |\hat R_{EP}\rangle=0$.
If desired, the spectral content in this asymptotic expressions can again be further evaluated by using Eq.~\eqref{eq:cpasym}.

Equation \eqref{eq:result2} implies that the additional states in the system provide  generally nonnegligible nonperturbative contributions to the phase rigidity. This is further illustrated by the examples in Sec.~\ref{sec:examples}.

\section{Constructive Approach}
\label{sec:proof2}
The derivation of the general result \eqref{eq:result0} in Sec.~\ref{sec:result} is closely connected to considerations in the eigenvectors-from-eigenvalues school of thought, while the analysis of the asymptotic behaviour near EPs in Sec.~\ref{sec:asymptotic} makes use of the separation of spectral data in the characteristic polynomial, and additional geometric data in the coefficients $g_i$. As laid out in the background Sec.~\ref{sec:background}, the determinantal minors appearing in these coefficients also allow to express spectral data.
Following the essence of these considerations further, we can therefore approach this asymptotic behaviour directly in terms of these minors.

For this we express the components of the non-normalized
eigenvectors freely and exactly as
\begin{align}
R_{i,k}&=(-1)^k M_{sk}(E_i\openone-H),\nonumber\\
L_{i,k}^*&=(-1)^k M_{kt}(E_i\openone-H).
\label{eq:explicit}
\end{align}
Away from an EP, we can choose the indices $s$ and $t$ arbitrarily, while at an EP these expressions apply to all choices of $s$ and $t$ that give a finite result, of which there is generically at least one.
The proof of this representation is simple---for any square matrix $A$ with $\mathrm{det}\,A=0$, $A \,\mathrm{adj}\,A=0$, so that each  column $|\mathbf{a}_s\rangle$ of  $\mathrm{adj}\,A$ provides a solution to the homogenous system of equations  $A |\mathbf{a}_s\rangle=0$. We here simply apply this  to the matrix $A=E_i\openone-H$, and obtain a right eigenvector as long as the result does not vanish. For the left eigenvector, we proceed analogously using the rows of $\mathrm{adj}\,A$, given that $\mathrm{adj}\,A\,A=0$.

For instance, using the triangular normal form $T$ of the truncated system \eqref{eq:h} and setting for convenience $E_{EP}=0$, while choosing   $s=1$, $t=n$, we can write
\begin{align}
R_{i,k}&\sim E_i^{n-k}\prod_{l=1}^{k-1}T_{l,l+1},
\\
L_{i,k}^*&\sim E_i^{k-1}\prod_{l=k}^{n-1}T_{l,l+1}.
\end{align}
It follows that $\langle R_i|R_i\rangle\sim\langle L_i|L_i\rangle\sim \xi^2 $,
 $|\langle L_i|R_i\rangle|\sim n|E_i^{n-1}|\xi$. Therefore,
$|r_i|\sim |nE_i^{n-1}|\xi/\xi^2$ again
recovers the result \eqref{eq:jan}, this time fully constructively.

We note that the general result \eqref{eq:result0} is significantly more compact than using Eq.~\eqref{eq:explicit}. On the other hand, the constructive approach gives direct access to the individual eigenvector components, hence,  captures their explicit geometry in a given basis. In the next section we utilize such additional insights to identify a general geometric feature hidden in these relations.

\section{Equipartition principle}
\label{sec:equi}

The intermediate steps of the original derivation of Eq.~\eqref{eq:jan}
in Ref.~\cite{Wiersig2023} involve a number of interesting perturbative relations in the space of the truncated system \eqref{eq:h}, most prominently
for the overlaps
\begin{equation}
\label{eq:inter}
\langle \hat L_i| \hat R_i \rangle\sim n\langle \hat L_{EP}|\hat R_i \rangle,
\end{equation}
where the subscript EP denotes quantities at the EP.
We did not use this relation, but note that it can be recovered from Eq.~\eqref{eq:imp}. For this, we simply take
\begin{align}
\langle \hat L_{EP} |A| \hat L_{EP}\rangle&\sim \frac{\langle \hat L_{EP}|\hat R_i\rangle}{\langle \hat L_i| \hat R_i\rangle} p'(E_i)
\nonumber\\
&=  M_{\hat L_{EP}\hat L_{EP}}((E_i-E_{EP})\openone-N).
\end{align}
Now, because of the reduced matrix size in the minors we find for the diagonal terms
$M_{\hat L_{EP}\hat L_{EP}}((E_i-E_{EP})\openone-N)\sim (E_i-E_{EP})^{n-1}\sim p'(E_i)/n$. The relation \eqref{eq:inter} for overlaps in the truncated system then follows directly.

\begin{figure}
  \centering
  \includegraphics[width=0.8\columnwidth]{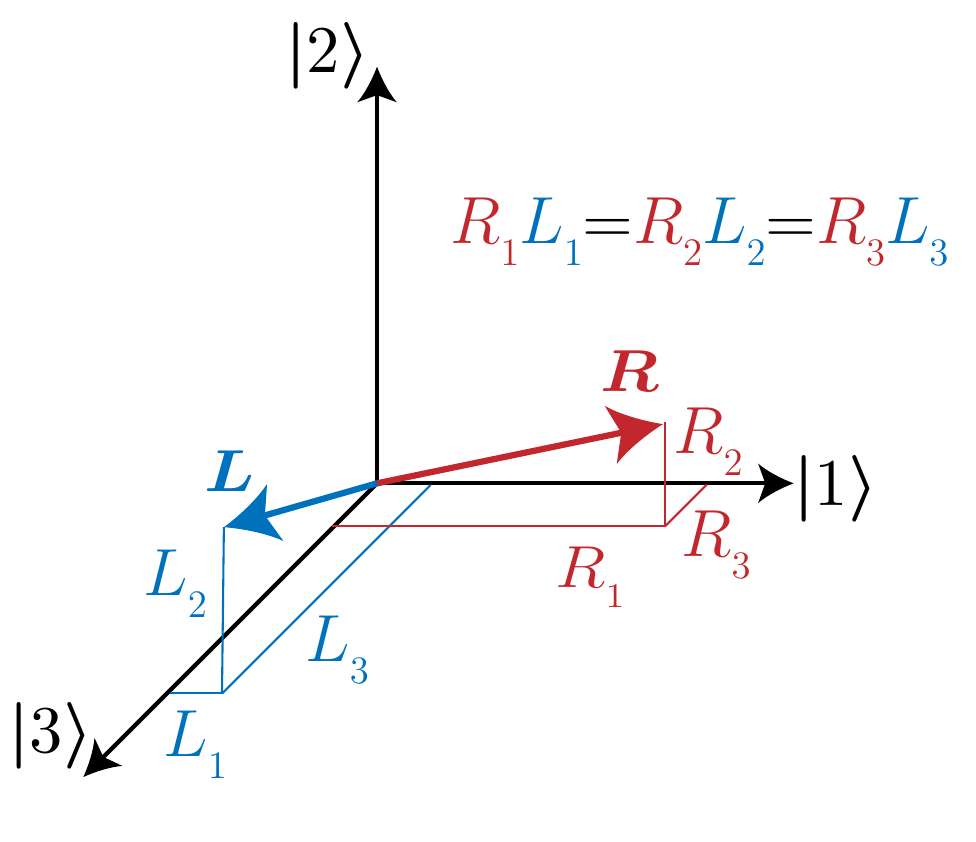}
  \caption{Illustration of the equipartition principle \eqref{eq:equip} in the truncated subspace  near an EP of order 3, for a pair of right and left eigenvectors $\mathbf{R}$, $\mathbf{L}$ with real components.}\label{fig1}
\end{figure}

Using our explicit expressions for the states, we can now concretize this relation to obtain geometric insights into the states near the EP. First, using the upper-triangular normal form $T$ of the still truncated system,
$M_{kk}((E_i-E_{EP})\openone-T)= (E_i-E_{EP})^{n-1}$ is identical for all $k$, and
Eq.~\eqref{eq:imp}  gives
\begin{equation}
\label{eq:equip}
R_{i,k}L_{i,k}^*=\frac{1}{n}\langle L_i|R_i\rangle,
\end{equation}
so that every individual term contributes exactly equally to this overlap.
Next, using the generality of expression \eqref{eq:imp}, we find that this still applies to the non-truncated case, as the further overlaps are all of a higher order.
Therefore, in the normal-form basis, each direction in the quasi-degenerate space provides exactly the same contribution  to the phase rigidity, as illustrated in Fig.~\ref{fig1}. This equipartition  principle provides a striking geometric reinterpretation of the factor $n$ appearing in the asymptotic result \eqref{eq:jan}, which still carries over to the general case, and severely restricts the eigenstate geometry near an EP.

\section{Illustrative examples}
\label{sec:examples}
We close this paper by providing two simple examples of systems close to an EP. The examples illuminate the role of additional states in the system, including for the case where these participate themselves in their own EP. This reveals the generally nonperturbative role of any additional states, which we condense into guidance for more general applications.

\subsection{Role of a spectrally isolated additional state}

As the first example, we consider a $3\times3$ Hamiltonian at an EP of order 2, written in upper triangular form as
\begin{equation}
H=\left(
    \begin{array}{ccc}
      0 & a & b \\
      0 & 0 & c \\
      0 & 0 & \Delta \\
    \end{array}
  \right).
\end{equation}
The states participating in the EP have eigenvalue $E_{EP}=E_{(1|2)}=0$, and the right and left eigenstate associated with it can be written as
\begin{align}
|R_{(1|2)}\rangle=\left(
    \begin{array}{c}
      1 \\
      0 \\
      0 \\
    \end{array}
  \right), \quad \langle L_{(1|2)}|=(0,\Delta,-c).
\end{align}
As required for a system that sits exactly at an EP,  these states are self-orthogonal, $\langle L_{(1|2)}|R_{(1|2)}\rangle=0$,
so that the phase rigidity $r_{(1|2)}=0$ vanishes.

Adding a small perturbation $\varepsilon H'$ to this Hamiltonian,
the eigenvalues participating in the EP become, in leading order, perturbed into
\begin{equation}
\label{eq:perteval1}
E_{1,2}\sim  \pm\sqrt{-\mathrm{tr}\,[H'\,\mathrm{adj}\,(-H)]/\Delta}\equiv \pm \delta_{(1|2)},
\end{equation}
 where we expressed this compactly using the
unperturbed adjugate matrix
\begin{align}
\mathrm{adj}\,(-H)&=\left(
    \begin{array}{cccc}
      0 & -a \Delta &  ac \\
      0 & 0 & 0\\
      0 & 0 & 0\\
    \end{array}
  \right).
  \end{align}
Utilizing the same matrix in Eq.~\eqref{eq:adef}, we evaluate
that in contrast to the phase rigidity itself, the coefficient
\begin{equation}
g_{(1|2)}=a\sqrt{|c|^2+|\Delta|^2}
\end{equation}
indeed remains finite at the EP.
Together with the asymptotic form $p'(E_{1,2})\sim -2 E_{1,2}\Delta$ of the characteristic polynomial,
we then find the asymptotics
\begin{equation}
\label{eq:ex1}
|r_{1,2}|\sim\frac{2|E_{1,2}|}{|a|}\frac{|\Delta|}{\sqrt{|c|^2+|\Delta|^2}}
\end{equation}
of the phase rigidity for each of the two eigenvalues that approach the EP.

These steps illuminate the passage from the general result \eqref{eq:result0} to its asymptotic form \eqref{eq:result2}.
In the final expression \eqref{eq:ex1}, the first factor $|2E_{1,2}/a|$ coincides with the approximation where the matrix would be truncated to the first and second row and column,
to which we then can apply Eq.~\eqref{eq:jan}, while the remaining factor captures the contribution from the overlap of the additional eigenstate with the quasi-degenerate subspace of the states near the EP.
Such additional states can therefore not be neglected.

Returning to the unperturbed system, the remaining eigenstate has eigenvalue $E_3=\Delta$, and
eigenvectors
\begin{align}
|R_{3}\rangle=\left(
    \begin{array}{c}
      (ac+b\Delta)/\Delta \\
      c \\
      \Delta \\
    \end{array}
  \right), \quad \langle L_{3}|=(0,0,1),
\end{align}
which from the outset have a finite overlap.
The phase rigidity quantifying this overlap,
\begin{equation}
r_3=\frac{|\Delta|^2}{\sqrt{|(ac+b\Delta)/\Delta|^2+|c|^2+|\Delta|^2}},
\end{equation}
indeed agrees exactly with the prediction of the general, nonperturbative, expression \eqref{eq:result0},
where we utilize $p'(\Delta)=\Delta^2$ and
\begin{equation}
\mathrm{adj}\,(\Delta\openone-H)=\left(
    \begin{array}{cccc}
     0 & 0 & ac+b\Delta\\
     0 & 0 &  \Delta c\\
     0 & 0  & \Delta^2\\
    \end{array}
   \right).
\end{equation}.

This illustrates how the framework developed in this work  applies to evaluate the phase rigidity both of quasidegenerate and isolated states in the system, and that in each case the result involves nonperturbative terms from the overlap of their respective subspaces.

\subsection{Interplay of exceptional points}
\label{sec:example}
As a second example, let us
consider a system of four states, which pairwise participate in two EPs.
We again utilize Schur's unitary triangularization theorem to study this system in its upper-triangular form,
\begin{equation}
H=\left(
    \begin{array}{cccc}
      0 & a_1 & a_2 & a_3\\
      0 & 0 & b_1 &  b_2\\
      0 & 0 & \Delta  & c_1\\
      0 & 0 & 0 & \Delta \\
    \end{array}
  \right),
\end{equation}
where one EP has a vanishing eigenvalue $E_1=E_2\equiv E_{(1|2)}=0$, while the other has the eigenvalue $E_3=E_4\equiv E_{(3|4)}=\Delta$.
Because of the defectiveness, each of these algebraically degenerate eigenvalues has only one right-left eigenvector pair,
\begin{align}
|R_{(1|2)}\rangle=\left(
    \begin{array}{c}
      1 \\
      0 \\
      0 \\
      0 \\
    \end{array}
  \right), \quad \langle L_{(1|2)}|=(0,\Delta,-b_1,(b_1c_1-b_2\Delta)/\Delta),
\\
|R_{(3|4)}\rangle=\left(
    \begin{array}{c}
      (a_1b_1+a_2\Delta)/\Delta\\
      b_1 \\
      \Delta \\
      0 \\
    \end{array}
  \right), \quad \langle L_{(3|4)}|=(0,0,0,1).
\end{align}
The corresponding adjugate matrices are
\begin{align}
\mathrm{adj}\,(-H)&=\left(
    \begin{array}{cccc}
      0 & a_1\Delta^2 & -a_1 b_1 \Delta & a_1(b_1c_1- b_2\Delta)\\
      0 & 0 & 0 &  0\\
      0 & 0 & 0  & 0\\
      0 & 0 & 0 & 0 \\
    \end{array}
  \right)
  ,
  \\
  \mathrm{adj}\,(\Delta\openone-H)&=\left(
    \begin{array}{cccc}
      0 & 0 & 0 & (a_1b_1+a_2\Delta)c_1\\
      0 & 0 & 0 &  \Delta b_1 c_1\\
      0 & 0 & 0  & \Delta^2 c_1\\
      0 & 0 & 0 & 0 \\
    \end{array}
   \right)
   .
\end{align}

Therefore, while at the EP the phase rigidity $r_i$ of the participating eigenvalues again vanishes,
the coefficients
\begin{align}
  g_{(1|2)}&= a_1 \Delta \sqrt{|\Delta^2|+|b_1|^2+|b_1c_1- b_2\Delta|^2/|\Delta|^2},\\
  g_{(3|4)} &=c_1\Delta\sqrt{|\Delta^2|+|b_1|^2+|a_1b_1+ a_2\Delta|^2/|\Delta|^2}
\end{align}
are once more both finite.

Adding a small perturbation $\varepsilon H'$, the perturbative eigenvalues can be  compactly expressed as
\begin{align}
E_{1,2}&\sim\pm \sqrt{ \varepsilon\mathrm{tr}\,[H'\,\mathrm{adj}\,(-H)]}/\Delta\equiv \pm \delta_{(1|2)} ,\nonumber\\
E_{3,4}&\sim\Delta\pm\sqrt{\varepsilon\mathrm{tr}\,[H'\,\mathrm{A}\,(\Delta-H)]}/\Delta\equiv \Delta\pm\delta_{(3|4)}.
\label{eq:perteval2}
\end{align}

Combining all these result in our asymptotic expression \eqref{eq:result2},
we therefore obtain
\begin{align}
|r_{1,2}|\sim\frac{2|\delta_{(1|2)}|}{|a_1|}\frac{1}{\sqrt{1+|b_1|^2/|\Delta^2|+|b_1c_1- b_2\Delta|^2/|\Delta|^4}}
\end{align}
and
\begin{align}
|r_{3,4}|\sim
\frac{2|\delta_{(3|4)}|}{|c_1|}
\frac{1}{\sqrt{1+|b_1|^2/|\Delta^2|+|a_1b_1+ a_2\Delta|^2/|\Delta|^4}}.
\end{align}
In both cases, this again contains a factor as in Eq.~\eqref{eq:jan}, representing the truncation of the system to the 2-dimensional quasi-degenerate subspace, and a contribution characterizing the overlap with the remaining states in the system.

\subsection{Further guidance for applications}
We note that in both examples, the additional states in the system provide generally nonnegligible, nonperturbative contributions to the phase rigidity.
Furthermore, while the eigenvalue perturbations \eqref{eq:perteval1}, \eqref{eq:perteval2} also pick up contributions from outside the quasidegenerate subspace, both effects do not cancel.
Therefore, in general, truncations to the quasidegenerate subspace have to be treated with caution---their justification not only requires an analysis of the matrix elements of the perturbation, but also of the nonperturbative matrix elements of the adjugate matrix.
In practical applications, this can be further guided by the constructive approach in Sec.~\ref{sec:proof2}, which gives direct access to the eigenvector components in a given basis.
However, in all these  settings, our general expression \eqref{eq:result0} and the asymptotic form \eqref{eq:result2} can  be utilized to analyze these without any further restrictions.
In particular, within this formalism we do not encounter any further complications from the fact that the remaining states may themselves be close to an EP.

\section{Conclusions}
\label{sec:conclusions}
In summary, I provide an exact nonperturbative reformulation of the phase rigidity (or equivalently, the eigenvalue condition number and the Petermann factor), given by Eq.~\eqref{eq:result0}. This expression has three key merits---it is exact, it is well behaved near exceptional points where it provides the asymptotic form Eq. \eqref{eq:result2}, and it does not require truncation of the system to a quasi-degenerate subspace.

The expression therefore enjoys a wide range of applicability, including to systems with additional overlapping states that possibly participate in their own exceptional points.
The result also holds, e.g., when no eigenvalues are close to an exceptional points, or when such multiple exceptional points are brought close to each other so that the quasi-degenerate eigenvalue clouds become intermingled.
This then enables the careful and precise analysis of effectively non-Hermitian systems, which is essential for the evaluation of their intriguingly modified sensitivity to static and dynamic perturbations.

The results also demonstrate that common truncations to quasi-degenerate subspaces have to be treated with caution. These are only admissible if other states in the system do not overlap with these spaces, which can only be asserted using additional arguments. The formalism described here identifies particular elements of the adjugate matrix as crucial for the justification of such truncations.

This realization should motivate the design of systems where new functionality arises from the combination of several separate exceptional points.
A particular interesting setting to explore such new enhancement effects are active systems, in which inhomogenous gain significantly changes the mode-nonorthogonality, while causality constraints of passive systems  \cite{Wiersig2019,Schomerus2022,Wiersig2022} no longer apply. This also includes nonreciprocal variants of these systems, which support directed amplification and sensing \cite{Sch20,Budich2020,Wanjura2020,McDonald2020}.

\begin{acknowledgments}
I gratefully thank Jan Wiersig for informing me about his work \cite{Wiersig2023}, as well as for insightful discussions, and Carlo Beenakker for sharing with me the insights of Ref.~\cite{denton2022eigenvectors} during an inspirational visit to Leiden.
The author acknowledges funding by EPSRC via Program Grant No. EP/N031776/1.
No new data were created during this study.
\end{acknowledgments}


\begin{thebibliography}{37}%
\makeatletter
\providecommand \@ifxundefined [1]{%
 \@ifx{#1\undefined}
}%
\providecommand \@ifnum [1]{%
 \ifnum #1\expandafter \@firstoftwo
 \else \expandafter \@secondoftwo
 \fi
}%
\providecommand \@ifx [1]{%
 \ifx #1\expandafter \@firstoftwo
 \else \expandafter \@secondoftwo
 \fi
}%
\providecommand \natexlab [1]{#1}%
\providecommand \enquote  [1]{``#1''}%
\providecommand \bibnamefont  [1]{#1}%
\providecommand \bibfnamefont [1]{#1}%
\providecommand \citenamefont [1]{#1}%
\providecommand \href@noop [0]{\@secondoftwo}%
\providecommand \href [0]{\begingroup \@sanitize@url \@href}%
\providecommand \@href[1]{\@@startlink{#1}\@@href}%
\providecommand \@@href[1]{\endgroup#1\@@endlink}%
\providecommand \@sanitize@url [0]{\catcode `\\12\catcode `\$12\catcode
  `\&12\catcode `\#12\catcode `\^12\catcode `\_12\catcode `\%12\relax}%
\providecommand \@@startlink[1]{}%
\providecommand \@@endlink[0]{}%
\providecommand \url  [0]{\begingroup\@sanitize@url \@url }%
\providecommand \@url [1]{\endgroup\@href {#1}{\urlprefix }}%
\providecommand \urlprefix  [0]{URL }%
\providecommand \Eprint [0]{\href }%
\providecommand \doibase [0]{https://doi.org/}%
\providecommand \selectlanguage [0]{\@gobble}%
\providecommand \bibinfo  [0]{\@secondoftwo}%
\providecommand \bibfield  [0]{\@secondoftwo}%
\providecommand \translation [1]{[#1]}%
\providecommand \BibitemOpen [0]{}%
\providecommand \bibitemStop [0]{}%
\providecommand \bibitemNoStop [0]{.\EOS\space}%
\providecommand \EOS [0]{\spacefactor3000\relax}%
\providecommand \BibitemShut  [1]{\csname bibitem#1\endcsname}%
\let\auto@bib@innerbib\@empty
\bibitem [{\citenamefont {Moiseyev}(2011)}]{moiseyev2011non}%
  \BibitemOpen
  \bibfield  {author} {\bibinfo {author} {\bibfnamefont {N.}~\bibnamefont
  {Moiseyev}},\ }\href@noop {} {\emph {\bibinfo {title} {Non-{Hermitian}
  quantum mechanics}}}\ (\bibinfo  {publisher} {Cambridge University Press},\
  \bibinfo {year} {2011})\BibitemShut {NoStop}%
\bibitem [{\citenamefont {Ashida}\ \emph {et~al.}(2020)\citenamefont {Ashida},
  \citenamefont {Gong},\ and\ \citenamefont {Ueda}}]{ashida2020}%
  \BibitemOpen
  \bibfield  {author} {\bibinfo {author} {\bibfnamefont {Y.}~\bibnamefont
  {Ashida}}, \bibinfo {author} {\bibfnamefont {Z.}~\bibnamefont {Gong}},\ and\
  \bibinfo {author} {\bibfnamefont {M.}~\bibnamefont {Ueda}},\ }\bibfield
  {title} {\bibinfo {title} {Non-{Hermitian} physics},\ }\href
  {https://doi.org/10.1080/00018732.2021.1876991} {\bibfield  {journal}
  {\bibinfo  {journal} {Adv. Phys.}\ }\textbf {\bibinfo {volume} {69}},\
  \bibinfo {pages} {249} (\bibinfo {year} {2020})}\BibitemShut {NoStop}%
\bibitem [{\citenamefont {Kato}(2013)}]{kato2013perturbation}%
  \BibitemOpen
  \bibfield  {author} {\bibinfo {author} {\bibfnamefont {T.}~\bibnamefont
  {Kato}},\ }\href@noop {} {\emph {\bibinfo {title} {Perturbation theory for
  linear operators}}}\ (\bibinfo  {publisher} {Springer Science \& Business
  Media},\ \bibinfo {year} {2013})\BibitemShut {NoStop}%
\bibitem [{\citenamefont {Heiss}(2000)}]{Heiss2000}%
  \BibitemOpen
  \bibfield  {author} {\bibinfo {author} {\bibfnamefont {W.~D.}\ \bibnamefont
  {Heiss}},\ }\bibfield  {title} {\bibinfo {title} {Repulsion of resonance
  states and exceptional points},\ }\href
  {https://doi.org/10.1103/PhysRevE.61.929} {\bibfield  {journal} {\bibinfo
  {journal} {Phys. Rev. E}\ }\textbf {\bibinfo {volume} {61}},\ \bibinfo
  {pages} {929} (\bibinfo {year} {2000})}\BibitemShut {NoStop}%
\bibitem [{\citenamefont {Berry}(2004)}]{Berry2004}%
  \BibitemOpen
  \bibfield  {author} {\bibinfo {author} {\bibfnamefont {M.~V.}\ \bibnamefont
  {Berry}},\ }\bibfield  {title} {\bibinfo {title} {Physics of nonhermitian
  degeneracies},\ }\href {https://doi.org/10.1023/B:CJOP.0000044002.05657.04}
  {\bibfield  {journal} {\bibinfo  {journal} {Czech. J. Physics}\ }\textbf
  {\bibinfo {volume} {54}},\ \bibinfo {pages} {1039} (\bibinfo {year}
  {2004})}\BibitemShut {NoStop}%
\bibitem [{\citenamefont {Heiss}(2004)}]{Heiss2004}%
  \BibitemOpen
  \bibfield  {author} {\bibinfo {author} {\bibfnamefont {W.~D.}\ \bibnamefont
  {Heiss}},\ }\bibfield  {title} {\bibinfo {title} {Exceptional points of
  non-{Hermitian} operators},\ }\href
  {https://doi.org/10.1088/0305-4470/37/6/034} {\bibfield  {journal} {\bibinfo
  {journal} {J. Phys. A}\ }\textbf {\bibinfo {volume} {37}},\ \bibinfo {pages}
  {2455} (\bibinfo {year} {2004})}\BibitemShut {NoStop}%
\bibitem [{\citenamefont {Wiersig}(2014)}]{Wie14}%
  \BibitemOpen
  \bibfield  {author} {\bibinfo {author} {\bibfnamefont {J.}~\bibnamefont
  {Wiersig}},\ }\bibfield  {title} {\bibinfo {title} {Enhancing the sensitivity
  of frequency and energy splitting detection by using exceptional points:
  Application to microcavity sensors for single-particle detection},\ }\href
  {https://doi.org/10.1103/PhysRevLett.112.203901} {\bibfield  {journal}
  {\bibinfo  {journal} {Phys. Rev. Lett.}\ }\textbf {\bibinfo {volume} {112}},\
  \bibinfo {pages} {203901} (\bibinfo {year} {2014})}\BibitemShut {NoStop}%
\bibitem [{\citenamefont {Wiersig}(2020)}]{Wiersig20}%
  \BibitemOpen
  \bibfield  {author} {\bibinfo {author} {\bibfnamefont {J.}~\bibnamefont
  {Wiersig}},\ }\bibfield  {title} {\bibinfo {title} {Review of exceptional
  point-based sensors},\ }\href {https://doi.org/10.1364/PRJ.396115} {\bibfield
   {journal} {\bibinfo  {journal} {Photon. Res.}\ }\textbf {\bibinfo {volume}
  {8}},\ \bibinfo {pages} {1457} (\bibinfo {year} {2020})}\BibitemShut
  {NoStop}%
\bibitem [{\citenamefont {Schomerus}(2020)}]{Sch20}%
  \BibitemOpen
  \bibfield  {author} {\bibinfo {author} {\bibfnamefont {H.}~\bibnamefont
  {Schomerus}},\ }\bibfield  {title} {\bibinfo {title} {Nonreciprocal response
  theory of non-{Hermitian} mechanical metamaterials: Response phase transition
  from the skin effect of zero modes},\ }\href
  {https://doi.org/10.1103/PhysRevResearch.2.013058} {\bibfield  {journal}
  {\bibinfo  {journal} {Phys. Rev. Research}\ }\textbf {\bibinfo {volume}
  {2}},\ \bibinfo {pages} {013058} (\bibinfo {year} {2020})}\BibitemShut
  {NoStop}%
\bibitem [{\citenamefont {Hashemi}\ \emph {et~al.}(2022)\citenamefont
  {Hashemi}, \citenamefont {Busch}, \citenamefont {Christodoulides},
  \citenamefont {Ozdemir},\ and\ \citenamefont {El-Ganainy}}]{Hashemi2022}%
  \BibitemOpen
  \bibfield  {author} {\bibinfo {author} {\bibfnamefont {A.}~\bibnamefont
  {Hashemi}}, \bibinfo {author} {\bibfnamefont {K.}~\bibnamefont {Busch}},
  \bibinfo {author} {\bibfnamefont {D.~N.}\ \bibnamefont {Christodoulides}},
  \bibinfo {author} {\bibfnamefont {S.~K.}\ \bibnamefont {Ozdemir}},\ and\
  \bibinfo {author} {\bibfnamefont {R.}~\bibnamefont {El-Ganainy}},\ }\bibfield
   {title} {\bibinfo {title} {Linear response theory of open systems with
  exceptional points},\ }\href {https://doi.org/10.1038/s41467-022-30715-8}
  {\bibfield  {journal} {\bibinfo  {journal} {Nat. Commun.}\ }\textbf {\bibinfo
  {volume} {13}},\ \bibinfo {pages} {3281} (\bibinfo {year}
  {2022})}\BibitemShut {NoStop}%
\bibitem [{\citenamefont {Budich}\ and\ \citenamefont
  {Bergholtz}(2020)}]{Budich2020}%
  \BibitemOpen
  \bibfield  {author} {\bibinfo {author} {\bibfnamefont {J.~C.}\ \bibnamefont
  {Budich}}\ and\ \bibinfo {author} {\bibfnamefont {E.~J.}\ \bibnamefont
  {Bergholtz}},\ }\bibfield  {title} {\bibinfo {title} {Non-{Hermitian}
  topological sensors},\ }\href
  {https://doi.org/10.1103/PhysRevLett.125.180403} {\bibfield  {journal}
  {\bibinfo  {journal} {Phys. Rev. Lett.}\ }\textbf {\bibinfo {volume} {125}},\
  \bibinfo {pages} {180403} (\bibinfo {year} {2020})}\BibitemShut {NoStop}%
\bibitem [{\citenamefont {van Langen}\ \emph {et~al.}(1997)\citenamefont {van
  Langen}, \citenamefont {Brouwer},\ and\ \citenamefont
  {Beenakker}}]{vanLangen1997}%
  \BibitemOpen
  \bibfield  {author} {\bibinfo {author} {\bibfnamefont {S.~A.}\ \bibnamefont
  {van Langen}}, \bibinfo {author} {\bibfnamefont {P.~W.}\ \bibnamefont
  {Brouwer}},\ and\ \bibinfo {author} {\bibfnamefont {C.~W.~J.}\ \bibnamefont
  {Beenakker}},\ }\bibfield  {title} {\bibinfo {title} {Fluctuating phase
  rigidity for a quantum chaotic system with partially broken time-reversal
  symmetry},\ }\href {https://doi.org/10.1103/PhysRevE.55.R1} {\bibfield
  {journal} {\bibinfo  {journal} {Phys. Rev. E}\ }\textbf {\bibinfo {volume}
  {55}},\ \bibinfo {pages} {R1} (\bibinfo {year} {1997})}\BibitemShut {NoStop}%
\bibitem [{\citenamefont {Petermann}(1979)}]{Petermann79}%
  \BibitemOpen
  \bibfield  {author} {\bibinfo {author} {\bibfnamefont {K.}~\bibnamefont
  {Petermann}},\ }\bibfield  {title} {\bibinfo {title} {Calculated spontaneous
  emission factor for double-heterostructure injection lasers with gain-induced
  waveguiding},\ }\href {https://doi.org/10.1109/JQE.1979.1070064} {\bibfield
  {journal} {\bibinfo  {journal} {IEEE J. Quantum Electron.}\ }\textbf
  {\bibinfo {volume} {15}},\ \bibinfo {pages} {566} (\bibinfo {year}
  {1979})}\BibitemShut {NoStop}%
\bibitem [{\citenamefont {Siegman}(1989)}]{Siegman89}%
  \BibitemOpen
  \bibfield  {author} {\bibinfo {author} {\bibfnamefont {A.~E.}\ \bibnamefont
  {Siegman}},\ }\bibfield  {title} {\bibinfo {title} {Excess spontaneous
  emission in non-{Hermitian} optical systems. {I.} {Laser} amplifiers},\
  }\href {https://doi.org/10.1103/PhysRevA.39.1253} {\bibfield  {journal}
  {\bibinfo  {journal} {Phys. Rev. A}\ }\textbf {\bibinfo {volume} {39}},\
  \bibinfo {pages} {1253} (\bibinfo {year} {1989})}\BibitemShut {NoStop}%
\bibitem [{\citenamefont {Patra}\ \emph {et~al.}(2000)\citenamefont {Patra},
  \citenamefont {Schomerus},\ and\ \citenamefont {Beenakker}}]{Patra2000}%
  \BibitemOpen
  \bibfield  {author} {\bibinfo {author} {\bibfnamefont {M.}~\bibnamefont
  {Patra}}, \bibinfo {author} {\bibfnamefont {H.}~\bibnamefont {Schomerus}},\
  and\ \bibinfo {author} {\bibfnamefont {C.~W.~J.}\ \bibnamefont {Beenakker}},\
  }\bibfield  {title} {\bibinfo {title} {Quantum-limited linewidth of a chaotic
  laser cavity},\ }\href {https://doi.org/10.1103/PhysRevA.61.023810}
  {\bibfield  {journal} {\bibinfo  {journal} {Phys. Rev. A}\ }\textbf {\bibinfo
  {volume} {61}},\ \bibinfo {pages} {023810} (\bibinfo {year}
  {2000})}\BibitemShut {NoStop}%
\bibitem [{\citenamefont {Berry}(2003)}]{Berry2003}%
  \BibitemOpen
  \bibfield  {author} {\bibinfo {author} {\bibfnamefont {M.~V.}\ \bibnamefont
  {Berry}},\ }\bibfield  {title} {\bibinfo {title} {Mode degeneracies and the
  {Petermann} excess-noise factor for unstable lasers},\ }\href
  {https://doi.org/10.1080/09500340308234532} {\bibfield  {journal} {\bibinfo
  {journal} {Journal of Modern Optics}\ }\textbf {\bibinfo {volume} {50}},\
  \bibinfo {pages} {63} (\bibinfo {year} {2003})}\BibitemShut {NoStop}%
\bibitem [{\citenamefont {Yoo}\ \emph {et~al.}(2011)\citenamefont {Yoo},
  \citenamefont {Sim},\ and\ \citenamefont {Schomerus}}]{Yoo11}%
  \BibitemOpen
  \bibfield  {author} {\bibinfo {author} {\bibfnamefont {G.}~\bibnamefont
  {Yoo}}, \bibinfo {author} {\bibfnamefont {H.-S.}\ \bibnamefont {Sim}},\ and\
  \bibinfo {author} {\bibfnamefont {H.}~\bibnamefont {Schomerus}},\ }\bibfield
  {title} {\bibinfo {title} {Quantum noise and mode nonorthogonality in
  non-{Hermitian} {PT}-symmetric optical resonators},\ }\href
  {https://doi.org/10.1103/PhysRevA.84.063833} {\bibfield  {journal} {\bibinfo
  {journal} {Phys. Rev. A}\ }\textbf {\bibinfo {volume} {84}},\ \bibinfo
  {pages} {063833} (\bibinfo {year} {2011})}\BibitemShut {NoStop}%
\bibitem [{\citenamefont {Simonson}\ \emph {et~al.}(2022)\citenamefont
  {Simonson}, \citenamefont {Ozdemir}, \citenamefont {Eisfeld}, \citenamefont
  {Metelmann},\ and\ \citenamefont {El-Ganainy}}]{Simonson22}%
  \BibitemOpen
  \bibfield  {author} {\bibinfo {author} {\bibfnamefont {L.}~\bibnamefont
  {Simonson}}, \bibinfo {author} {\bibfnamefont {S.~K.}\ \bibnamefont
  {Ozdemir}}, \bibinfo {author} {\bibfnamefont {A.}~\bibnamefont {Eisfeld}},
  \bibinfo {author} {\bibfnamefont {A.}~\bibnamefont {Metelmann}},\ and\
  \bibinfo {author} {\bibfnamefont {R.}~\bibnamefont {El-Ganainy}},\ }\bibfield
   {title} {\bibinfo {title} {Nonuniversality of quantum noise in optical
  amplifiers operating at exceptional points},\ }\href
  {https://doi.org/10.1103/PhysRevResearch.4.033226} {\bibfield  {journal}
  {\bibinfo  {journal} {Phys. Rev. Res.}\ }\textbf {\bibinfo {volume} {4}},\
  \bibinfo {pages} {033226} (\bibinfo {year} {2022})}\BibitemShut {NoStop}%
\bibitem [{\citenamefont {van Eijkelenborg}\ \emph {et~al.}(1996)\citenamefont
  {van Eijkelenborg}, \citenamefont {Lindberg}, \citenamefont {Thijssen},\ and\
  \citenamefont {Woerdman}}]{vanEijkelenborg1996}%
  \BibitemOpen
  \bibfield  {author} {\bibinfo {author} {\bibfnamefont {M.~A.}\ \bibnamefont
  {van Eijkelenborg}}, \bibinfo {author} {\bibfnamefont {A.~M.}\ \bibnamefont
  {Lindberg}}, \bibinfo {author} {\bibfnamefont {M.~S.}\ \bibnamefont
  {Thijssen}},\ and\ \bibinfo {author} {\bibfnamefont {J.~P.}\ \bibnamefont
  {Woerdman}},\ }\bibfield  {title} {\bibinfo {title} {Resonance of quantum
  noise in an unstable cavity laser},\ }\href
  {https://doi.org/10.1103/PhysRevLett.77.4314} {\bibfield  {journal} {\bibinfo
   {journal} {Phys. Rev. Lett.}\ }\textbf {\bibinfo {volume} {77}},\ \bibinfo
  {pages} {4314} (\bibinfo {year} {1996})}\BibitemShut {NoStop}%
\bibitem [{\citenamefont {Cheng}\ \emph {et~al.}(1996)\citenamefont {Cheng},
  \citenamefont {Fanning},\ and\ \citenamefont {Siegman}}]{Cheng1996}%
  \BibitemOpen
  \bibfield  {author} {\bibinfo {author} {\bibfnamefont {Y.-J.}\ \bibnamefont
  {Cheng}}, \bibinfo {author} {\bibfnamefont {C.~G.}\ \bibnamefont {Fanning}},\
  and\ \bibinfo {author} {\bibfnamefont {A.~E.}\ \bibnamefont {Siegman}},\
  }\bibfield  {title} {\bibinfo {title} {Experimental observation of a large
  excess quantum noise factor in the linewidth of a laser oscillator having
  nonorthogonal modes},\ }\href {https://doi.org/10.1103/PhysRevLett.77.627}
  {\bibfield  {journal} {\bibinfo  {journal} {Phys. Rev. Lett.}\ }\textbf
  {\bibinfo {volume} {77}},\ \bibinfo {pages} {627} (\bibinfo {year}
  {1996})}\BibitemShut {NoStop}%
\bibitem [{\citenamefont {Takata}\ \emph {et~al.}(2021)\citenamefont {Takata},
  \citenamefont {Nozaki}, \citenamefont {Kuramochi}, \citenamefont {Matsuo},
  \citenamefont {Takeda}, \citenamefont {Fujii}, \citenamefont {Kita},
  \citenamefont {Shinya},\ and\ \citenamefont {Notomi}}]{Takata21}%
  \BibitemOpen
  \bibfield  {author} {\bibinfo {author} {\bibfnamefont {K.}~\bibnamefont
  {Takata}}, \bibinfo {author} {\bibfnamefont {K.}~\bibnamefont {Nozaki}},
  \bibinfo {author} {\bibfnamefont {E.}~\bibnamefont {Kuramochi}}, \bibinfo
  {author} {\bibfnamefont {S.}~\bibnamefont {Matsuo}}, \bibinfo {author}
  {\bibfnamefont {K.}~\bibnamefont {Takeda}}, \bibinfo {author} {\bibfnamefont
  {T.}~\bibnamefont {Fujii}}, \bibinfo {author} {\bibfnamefont
  {S.}~\bibnamefont {Kita}}, \bibinfo {author} {\bibfnamefont {A.}~\bibnamefont
  {Shinya}},\ and\ \bibinfo {author} {\bibfnamefont {M.}~\bibnamefont
  {Notomi}},\ }\bibfield  {title} {\bibinfo {title} {Observing exceptional
  point degeneracy of radiation with electrically pumped photonic crystal
  coupled-nanocavity lasers},\ }\href {https://doi.org/10.1364/OPTICA.412596}
  {\bibfield  {journal} {\bibinfo  {journal} {Optica}\ }\textbf {\bibinfo
  {volume} {8}},\ \bibinfo {pages} {184} (\bibinfo {year} {2021})}\BibitemShut
  {NoStop}%
\bibitem [{\citenamefont {Zhang}\ \emph {et~al.}(2018)\citenamefont {Zhang},
  \citenamefont {Peng}, \citenamefont {{\"O}zdemir}, \citenamefont {Pichler},
  \citenamefont {Krimer}, \citenamefont {Zhao}, \citenamefont {Nori},
  \citenamefont {Liu}, \citenamefont {Rotter},\ and\ \citenamefont
  {Yang}}]{Zhang2018}%
  \BibitemOpen
  \bibfield  {author} {\bibinfo {author} {\bibfnamefont {J.}~\bibnamefont
  {Zhang}}, \bibinfo {author} {\bibfnamefont {B.}~\bibnamefont {Peng}},
  \bibinfo {author} {\bibfnamefont {{\c{S}}.~K.}\ \bibnamefont {{\"O}zdemir}},
  \bibinfo {author} {\bibfnamefont {K.}~\bibnamefont {Pichler}}, \bibinfo
  {author} {\bibfnamefont {D.~O.}\ \bibnamefont {Krimer}}, \bibinfo {author}
  {\bibfnamefont {G.}~\bibnamefont {Zhao}}, \bibinfo {author} {\bibfnamefont
  {F.}~\bibnamefont {Nori}}, \bibinfo {author} {\bibfnamefont {Y.-x.}\
  \bibnamefont {Liu}}, \bibinfo {author} {\bibfnamefont {S.}~\bibnamefont
  {Rotter}},\ and\ \bibinfo {author} {\bibfnamefont {L.}~\bibnamefont {Yang}},\
  }\bibfield  {title} {\bibinfo {title} {A phonon laser operating at an
  exceptional point},\ }\href {https://doi.org/10.1038/s41566-018-0213-5}
  {\bibfield  {journal} {\bibinfo  {journal} {Nature Photonics}\ }\textbf
  {\bibinfo {volume} {12}},\ \bibinfo {pages} {479} (\bibinfo {year}
  {2018})}\BibitemShut {NoStop}%
\bibitem [{\citenamefont {Wang}\ \emph {et~al.}(2020)\citenamefont {Wang},
  \citenamefont {Lai}, \citenamefont {Yuan}, \citenamefont {Suh},\ and\
  \citenamefont {Vahala}}]{Wang2020}%
  \BibitemOpen
  \bibfield  {author} {\bibinfo {author} {\bibfnamefont {H.}~\bibnamefont
  {Wang}}, \bibinfo {author} {\bibfnamefont {Y.-H.}\ \bibnamefont {Lai}},
  \bibinfo {author} {\bibfnamefont {Z.}~\bibnamefont {Yuan}}, \bibinfo {author}
  {\bibfnamefont {M.-G.}\ \bibnamefont {Suh}},\ and\ \bibinfo {author}
  {\bibfnamefont {K.}~\bibnamefont {Vahala}},\ }\bibfield  {title} {\bibinfo
  {title} {Petermann-factor sensitivity limit near an exceptional point in a
  brillouin ring laser gyroscope},\ }\href
  {https://doi.org/10.1038/s41467-020-15341-6} {\bibfield  {journal} {\bibinfo
  {journal} {Nature Communications}\ }\textbf {\bibinfo {volume} {11}},\
  \bibinfo {pages} {1610} (\bibinfo {year} {2020})}\BibitemShut {NoStop}%
\bibitem [{\citenamefont {Wanjura}\ \emph {et~al.}(2020)\citenamefont
  {Wanjura}, \citenamefont {Brunelli},\ and\ \citenamefont
  {Nunnenkamp}}]{Wanjura2020}%
  \BibitemOpen
  \bibfield  {author} {\bibinfo {author} {\bibfnamefont {C.~C.}\ \bibnamefont
  {Wanjura}}, \bibinfo {author} {\bibfnamefont {M.}~\bibnamefont {Brunelli}},\
  and\ \bibinfo {author} {\bibfnamefont {A.}~\bibnamefont {Nunnenkamp}},\
  }\bibfield  {title} {\bibinfo {title} {Topological framework for directional
  amplification in driven-dissipative cavity arrays},\ }\href
  {https://doi.org/10.1038/s41467-020-16863-9} {\bibfield  {journal} {\bibinfo
  {journal} {Nat. Commun.}\ }\textbf {\bibinfo {volume} {11}},\ \bibinfo
  {pages} {3149} (\bibinfo {year} {2020})}\BibitemShut {NoStop}%
\bibitem [{\citenamefont {McDonald}\ and\ \citenamefont
  {Clerk}(2020)}]{McDonald2020}%
  \BibitemOpen
  \bibfield  {author} {\bibinfo {author} {\bibfnamefont {A.}~\bibnamefont
  {McDonald}}\ and\ \bibinfo {author} {\bibfnamefont {A.~A.}\ \bibnamefont
  {Clerk}},\ }\bibfield  {title} {\bibinfo {title} {Exponentially-enhanced
  quantum sensing with non-hermitian lattice dynamics},\ }\href
  {https://doi.org/10.1038/s41467-020-19090-4} {\bibfield  {journal} {\bibinfo
  {journal} {Nature Communications}\ }\textbf {\bibinfo {volume} {11}},\
  \bibinfo {pages} {5382} (\bibinfo {year} {2020})}\BibitemShut {NoStop}%
\bibitem [{\citenamefont {Doppler}\ \emph {et~al.}(2016)\citenamefont
  {Doppler}, \citenamefont {Mailybaev}, \citenamefont {B{\"o}hm}, \citenamefont
  {Kuhl}, \citenamefont {Girschik}, \citenamefont {Libisch}, \citenamefont
  {Milburn}, \citenamefont {Rabl}, \citenamefont {Moiseyev},\ and\
  \citenamefont {Rotter}}]{Doppler2016}%
  \BibitemOpen
  \bibfield  {author} {\bibinfo {author} {\bibfnamefont {J.}~\bibnamefont
  {Doppler}}, \bibinfo {author} {\bibfnamefont {A.~A.}\ \bibnamefont
  {Mailybaev}}, \bibinfo {author} {\bibfnamefont {J.}~\bibnamefont {B{\"o}hm}},
  \bibinfo {author} {\bibfnamefont {U.}~\bibnamefont {Kuhl}}, \bibinfo {author}
  {\bibfnamefont {A.}~\bibnamefont {Girschik}}, \bibinfo {author}
  {\bibfnamefont {F.}~\bibnamefont {Libisch}}, \bibinfo {author} {\bibfnamefont
  {T.~J.}\ \bibnamefont {Milburn}}, \bibinfo {author} {\bibfnamefont
  {P.}~\bibnamefont {Rabl}}, \bibinfo {author} {\bibfnamefont {N.}~\bibnamefont
  {Moiseyev}},\ and\ \bibinfo {author} {\bibfnamefont {S.}~\bibnamefont
  {Rotter}},\ }\bibfield  {title} {\bibinfo {title} {Dynamically encircling an
  exceptional point for asymmetric mode switching},\ }\href
  {https://doi.org/10.1038/nature18605} {\bibfield  {journal} {\bibinfo
  {journal} {Nature}\ }\textbf {\bibinfo {volume} {537}},\ \bibinfo {pages}
  {76} (\bibinfo {year} {2016})}\BibitemShut {NoStop}%
\bibitem [{\citenamefont {Bergholtz}\ \emph {et~al.}(2021)\citenamefont
  {Bergholtz}, \citenamefont {Budich},\ and\ \citenamefont {Kunst}}]{Ber19}%
  \BibitemOpen
  \bibfield  {author} {\bibinfo {author} {\bibfnamefont {E.~J.}\ \bibnamefont
  {Bergholtz}}, \bibinfo {author} {\bibfnamefont {J.~C.}\ \bibnamefont
  {Budich}},\ and\ \bibinfo {author} {\bibfnamefont {F.~K.}\ \bibnamefont
  {Kunst}},\ }\bibfield  {title} {\bibinfo {title} {Exceptional topology of
  non-{Hermitian} systems},\ }\href
  {https://doi.org/10.1103/RevModPhys.93.015005} {\bibfield  {journal}
  {\bibinfo  {journal} {Rev. Mod. Phys.}\ }\textbf {\bibinfo {volume} {93}},\
  \bibinfo {pages} {015005} (\bibinfo {year} {2021})}\BibitemShut {NoStop}%
\bibitem [{\citenamefont {Kawabata}\ \emph {et~al.}(2019)\citenamefont
  {Kawabata}, \citenamefont {Shiozaki}, \citenamefont {Ueda},\ and\
  \citenamefont {Sato}}]{Kaw19}%
  \BibitemOpen
  \bibfield  {author} {\bibinfo {author} {\bibfnamefont {K.}~\bibnamefont
  {Kawabata}}, \bibinfo {author} {\bibfnamefont {K.}~\bibnamefont {Shiozaki}},
  \bibinfo {author} {\bibfnamefont {M.}~\bibnamefont {Ueda}},\ and\ \bibinfo
  {author} {\bibfnamefont {M.}~\bibnamefont {Sato}},\ }\bibfield  {title}
  {\bibinfo {title} {Symmetry and topology in non-{Hermitian} physics},\ }\href
  {https://doi.org/10.1103/PhysRevX.9.041015} {\bibfield  {journal} {\bibinfo
  {journal} {Phys. Rev. X}\ }\textbf {\bibinfo {volume} {9}},\ \bibinfo {pages}
  {041015} (\bibinfo {year} {2019})}\BibitemShut {NoStop}%
\bibitem [{\citenamefont {Trefethen}(2005)}]{trefethen2005spectra}%
  \BibitemOpen
  \bibfield  {author} {\bibinfo {author} {\bibfnamefont {L.~N.}\ \bibnamefont
  {Trefethen}},\ }\bibfield  {title} {\bibinfo {title} {Spectra and
  pseudospectra: The behaviour of non-normal matrices and operators},\ }in\
  \href@noop {} {\emph {\bibinfo {booktitle} {The graduate student’s guide to
  numerical analysis’ 98: Lecture notes from the VIII EPSRC Summer School in
  Numerical Analysis}}}\ (\bibinfo  {publisher} {Springer},\ \bibinfo {year}
  {2005})\ pp.\ \bibinfo {pages} {217--250}\BibitemShut {NoStop}%
\bibitem [{\citenamefont {Chalker}\ and\ \citenamefont
  {Mehlig}(1998)}]{Chalker1998}%
  \BibitemOpen
  \bibfield  {author} {\bibinfo {author} {\bibfnamefont {J.~T.}\ \bibnamefont
  {Chalker}}\ and\ \bibinfo {author} {\bibfnamefont {B.}~\bibnamefont
  {Mehlig}},\ }\bibfield  {title} {\bibinfo {title} {Eigenvector statistics in
  non-hermitian random matrix ensembles},\ }\href
  {https://doi.org/10.1103/PhysRevLett.81.3367} {\bibfield  {journal} {\bibinfo
   {journal} {Phys. Rev. Lett.}\ }\textbf {\bibinfo {volume} {81}},\ \bibinfo
  {pages} {3367} (\bibinfo {year} {1998})}\BibitemShut {NoStop}%
\bibitem [{\citenamefont {Wiersig}(2019)}]{Wiersig2019}%
  \BibitemOpen
  \bibfield  {author} {\bibinfo {author} {\bibfnamefont {J.}~\bibnamefont
  {Wiersig}},\ }\bibfield  {title} {\bibinfo {title} {Nonorthogonality
  constraints in open quantum and wave systems},\ }\href
  {https://doi.org/10.1103/PhysRevResearch.1.033182} {\bibfield  {journal}
  {\bibinfo  {journal} {Phys. Rev. Res.}\ }\textbf {\bibinfo {volume} {1}},\
  \bibinfo {pages} {033182} (\bibinfo {year} {2019})}\BibitemShut {NoStop}%
\bibitem [{\citenamefont {Schomerus}(2022)}]{Schomerus2022}%
  \BibitemOpen
  \bibfield  {author} {\bibinfo {author} {\bibfnamefont {H.}~\bibnamefont
  {Schomerus}},\ }\bibfield  {title} {\bibinfo {title} {Fundamental constraints
  on the observability of non-{Hermitian} effects in passive systems},\ }\href
  {https://doi.org/10.1103/PhysRevA.106.063509} {\bibfield  {journal} {\bibinfo
   {journal} {Phys. Rev. A}\ }\textbf {\bibinfo {volume} {106}},\ \bibinfo
  {pages} {063509} (\bibinfo {year} {2022})}\BibitemShut {NoStop}%
\bibitem [{\citenamefont {Wiersig}(2022)}]{Wiersig2022}%
  \BibitemOpen
  \bibfield  {author} {\bibinfo {author} {\bibfnamefont {J.}~\bibnamefont
  {Wiersig}},\ }\bibfield  {title} {\bibinfo {title} {Distance between
  exceptional points and diabolic points and its implication for the response
  strength of non-{Hermitian} systems},\ }\href
  {https://doi.org/10.1103/PhysRevResearch.4.033179} {\bibfield  {journal}
  {\bibinfo  {journal} {Phys. Rev. Res.}\ }\textbf {\bibinfo {volume} {4}},\
  \bibinfo {pages} {033179} (\bibinfo {year} {2022})}\BibitemShut {NoStop}%
\bibitem [{\citenamefont {Denton}\ \emph {et~al.}(2022)\citenamefont {Denton},
  \citenamefont {Parke}, \citenamefont {Tao},\ and\ \citenamefont
  {Zhang}}]{denton2022eigenvectors}%
  \BibitemOpen
  \bibfield  {author} {\bibinfo {author} {\bibfnamefont {P.}~\bibnamefont
  {Denton}}, \bibinfo {author} {\bibfnamefont {S.}~\bibnamefont {Parke}},
  \bibinfo {author} {\bibfnamefont {T.}~\bibnamefont {Tao}},\ and\ \bibinfo
  {author} {\bibfnamefont {X.}~\bibnamefont {Zhang}},\ }\bibfield  {title}
  {\bibinfo {title} {Eigenvectors from eigenvalues: A survey of a basic
  identity in linear algebra},\ }\href@noop {} {\bibfield  {journal} {\bibinfo
  {journal} {Bull. Am. Math. Soc.}\ }\textbf {\bibinfo {volume} {59}},\
  \bibinfo {pages} {31} (\bibinfo {year} {2022})}\BibitemShut {NoStop}%
\bibitem [{\citenamefont {Wiersig}(2023)}]{Wiersig2023}%
  \BibitemOpen
  \bibfield  {author} {\bibinfo {author} {\bibfnamefont {J.}~\bibnamefont
  {Wiersig}},\ }\bibfield  {title} {\bibinfo {title} {Petermann factors and
  phase rigidities near exceptional points},\ }\href
  {https://doi.org/10.1103/PhysRevResearch.5.033042} {\bibfield  {journal}
  {\bibinfo  {journal} {Phys. Rev. Res.}\ }\textbf {\bibinfo {volume} {5}},\
  \bibinfo {pages} {033042} (\bibinfo {year} {2023})}\BibitemShut {NoStop}%
\bibitem [{Note1()}]{Note1}%
  \BibitemOpen
  \bibinfo {note} {Utilizing a similarity transformation, at an EP the system
  may further be brought into a Jordan normal form. However, this removes the
  geometric relations captured by $r_i$, whose physical content is fixed by the
  scalar product in the original Hilbert space.}\BibitemShut {Stop}%
\bibitem [{\citenamefont {Haake}\ \emph {et~al.}(1996)\citenamefont {Haake},
  \citenamefont {Kus}, \citenamefont {Sommers}, \citenamefont {Schomerus},\
  and\ \citenamefont {Zyczkowski}}]{Haake1996}%
  \BibitemOpen
  \bibfield  {author} {\bibinfo {author} {\bibfnamefont {F.}~\bibnamefont
  {Haake}}, \bibinfo {author} {\bibfnamefont {M.}~\bibnamefont {Kus}}, \bibinfo
  {author} {\bibfnamefont {H.-J.}\ \bibnamefont {Sommers}}, \bibinfo {author}
  {\bibfnamefont {H.}~\bibnamefont {Schomerus}},\ and\ \bibinfo {author}
  {\bibfnamefont {K.}~\bibnamefont {Zyczkowski}},\ }\bibfield  {title}
  {\bibinfo {title} {Secular determinants of random unitary matrices},\ }\href
  {https://doi.org/10.1088/0305-4470/29/13/029} {\bibfield  {journal} {\bibinfo
   {journal} {J. Phys. A}\ }\textbf {\bibinfo {volume} {29}},\ \bibinfo {pages}
  {3641} (\bibinfo {year} {1996})}\BibitemShut {NoStop}%
\end{thebibliography}
\end{document}